\documentclass[aps,prd,reprint,twocolumn,superscriptaddress,floatfix,preprintnumbers,nofootinbib]{revtex4-2}

\usepackage{graphicx,color,xcolor}
\usepackage{amsmath,amssymb,amsfonts,amsthm}
\usepackage{wasysym}
\usepackage[caption=false]{subfig}
\usepackage[colorlinks=True, citecolor=blue, urlcolor=blue, linkcolor=blue]{hyperref}
\usepackage{orcidlink}
\usepackage{ulem}
\usepackage{comment}
\usepackage{physics}
\usepackage{dsfont}

\newcommand{\AF}{\mathcal{F}_A}

\begin{document}

\title{The Magic Barrier: A Universal Feature of Quantum Thermalization}

\author{Jakob J.~Minar${}^{\orcidlink{0009-0008-5557-6919}}$}
\affiliation{Institut f\"ur Theoretische Physik, Universit\"at Regensburg, Regensburg, D-93040, Germany}

\author{Andreas Sch\"{a}fer${}^{\orcidlink{0000-0003-1171-0078}}$}
\affiliation{Institut f\"ur Theoretische Physik, Universit\"at Regensburg, Regensburg, D-93040, Germany}

\author{Leonhard Schmotzer${}^{\orcidlink{0009-0006-9307-072X}}$}
\affiliation{Institut f\"ur Theoretische Physik, Universit\"at Regensburg, Regensburg, D-93040, Germany}

\author{Clemens Seidl${}^{\orcidlink{0009-0006-9409-7553}}$}
\email{clemens.seidl@physik.uni-regensburg.de}
\affiliation{Institut f\"ur Theoretische Physik, Universit\"at Regensburg, Regensburg, D-93040, Germany}
\affiliation{Department of Physics and Arnold Sommerfeld Center for Theoretical Physics (ASC), Ludwig Maximilian University of Munich, 80333 M\"unchen, Germany}
\affiliation{Munich Center for Quantum Science and Technology (MCQST), 80799 M\"unchen, Germany}

\date{\today}

\begin{abstract}
    We show that the emergence of a ``magic barrier''~\cite{Ebner:2025pdm}, characterized by a transient maximum of nonstabilizerness during rapid entanglement growth, is a universal feature of thermalization within quantum ergodic theories, which is intimately connected to random matrix spectral properties. Starting from a pure product state, we analytically determine the time evolution of entanglement entropy, antiflatness, capacity of entanglement, and nonlocal trace distance magic under unitary dynamics generated by random matrix Hamiltonians drawn from all three Gaussian ensembles.
    We find that the transient maxima of antiflatness, capacity of entanglement, and nonlocal magic occur on the same characteristic timescale as the maximum rate of entanglement growth.
    These results establish that the magic barrier is a universal feature of RMT thermalization, suggesting universality across quantum chaotic systems. We conclude that quantitative simulations of thermalization in ergodic quantum theories inherently require quantum computation.
\end{abstract}

\maketitle

\section{Introduction}

Thermalization is the quantum analogue of ergodicity in classical systems \cite{Weidenmuller:2022wdc} and Random Matrix Theory (RMT) has clarified in great detail that many quantum systems show universal properties, which only depend on their fundamental symmetries. Recently we have shown that two rather different systems with RMT statistics, namely a SU(2) gauge theory plaquette chain and the mixed-field Ising model show the same type of ``magic barrier"~\cite{Ebner:2025pdm}, i.e. a barrier-like maximum in the time evolution of nonlocal nonstabilizerness during rapid entanglement growth.
Further investigations of e.g. random circuit dynamics~\cite{Aditya:2025yex}, random XXZ chains~\cite{Zhang:2026ucb}, Abelian $\mathbb{Z}_2$ and U(1) lattice gauge theories~\cite{Ebner:2026kwq} suggest that the appearance of such a barrier is indeed characteristic for thermalization of ergodic quantum many body systems.

In this work we demonstrate that the magic barrier is a universal feature of quantum-ergodic thermalization by calculating different entanglement and magic measures for an initial product state under random unitary time evolution in RMT.
The characteristic behavior is shown to be present at leading order for random spectral statistics of GOE, GUE and GSE, which denote the Gaussian orthogonal, unitary and symplectic ensemble, respectively.

Note that this does not exclude the possibility that thermalization of some states, e.g. quantum many-body scars, does not show this barrier, but for the vast majority of states it does. 

One can argue that it is really no great surprise that the thermalization of an increasingly entangled many-particle quantum system can only be completely described by a quantum computer.
On the other hand, the Gottesman-Knill theorem \cite{Aaronson:2004xuh} states that entanglement alone is insufficient to imply quantum advantage, and more precisely that qubit circuits using only Clifford gates cannot achieve (exponential) advantage compared to classical computers.
Further, it is well know that early low-entangled and final quasi-thermal time regimes can be efficiently treated classically. The crucial point is that our observation allows to study this fact quantitatively and thus identify the time regimes relevant for potential quantum advantage.

We concluded that a complete quantitative theoretical description of thermalization in a broad class of physical systems, i.e. those obeying RMT statistics, requires quantum computing.
Important realizations include strongly interacting many-body systems in condensed matter physics, atomic and molecular systems, non-Abelian gauge theories in nuclear and high-energy physics, and quantum chaotic systems relevant to black hole physics and holography.

RMT has developed over the years to a vast field of research and this is not the place to review this development. Some more relevant articles focused on thermalization and related aspects are \cite{Odavic:2024lqq,Rigol:2007juv,Srednicki:1994mfb, Iannotti:2026jvd,Cusumano:2026fal,Verde:2026hcg,Ebner:2026kwq}.
Naturally, the rapid improvement of quantum computing capability is of key importance for any discussion of the status of magic. Some typical references are
\cite{Daley:2022eja,Fauseweh:2024cow,Farrell:2023fgd,Ciavarella:2021nmj,Bauer:2022hpo}.

This discussion naturally raises the question of whether thermalization in quantum chromodynamics (QCD), the experimentally accessible theory of non-Abelian gauge interactions, also exhibits a ``magic barrier''. Ultrarelativistic heavy-ion collisions at the LHC provide a particularly compelling setting to investigate this question, as they produce strongly interacting matter that evolves as an approximately isolated quantum system on timescales many orders of magnitude shorter than those associated with particle detection.
While, to the best of our knowledge, there is no formal proof that non-Abelian gauge theories obey RMT statistics, strong numerical evidence suggests that this is indeed the case (see e.g. 
\cite{Verbaarschot:2000dy,Berbenni-Bitsch:1997zmi,Gockeler:1998jj,Gockeler:2001du,Pandey:2024goi}).
Remarkably, quantum magic has already been experimentally observed at the LHC in top quark-antiquark ($t\bar{t}$) production in proton-proton collisions~\cite{ATLAS:2023fsd,White:2024nuc}. While these measurements establish the presence of non-stabilizerness in high-energy scattering processes, our results suggest that the generation of magic may also be a generic feature of the far-from-equilibrium dynamics of strongly interacting quantum systems. In particular, they motivate the search for a transient ``magic barrier'' during the thermalization of QCD matter produced in ultrarelativistic heavy-ion collisions.

Further, the relationship between RMT and quantum gravity has attracted considerable attention, particularly in the context of black hole physics, where RMT provides a framework for understanding universal features of quantum chaotic spectra and their gravitational interpretation \cite{Jafferis:2025jle,Stanford:2019vob}.
At the same time, recent investigations have explored the role of nonstabilizerness in gravitational systems and holography, suggesting that it may provide a complementary perspective on the quantum information structure of spacetime \cite{Cao:2024nrx,White:2020zoz,Baiguera:2025dkc}.
Our results establish the connection between these two lines by demonstrating how a transient enhancement of nonstabilizerness emerges naturally within RMT descriptions of thermalization.

Taken together, these observations suggest a deeper connection between quantum chaos, thermalization, and quantum magic. Although entanglement and magic represent distinct quantum resources, their interplay during many-body dynamics remains to be fully understood \cite{Tirrito:2023fnw,Turkeshi:2024pnj,Tirrito:2024kts,Aditya:2026qna}. 
Previous investigations of random-state ensembles and random quantum circuits have already provided important insights into this relationship \cite{Xiao:2026tku,Zhang:2024fyp,Turkeshi:2023lqu,Chen:2022yza}.
However, whether the transient emergence of a ``magic barrier'' can be derived directly within random-matrix Hamiltonian dynamics remained an open question.
The Eigenstate Thermalization Hypothesis (ETH) \cite{DAlessio:2015qtq} and its relation to RMT \cite{Wang:2021mtp,Wang:2026hde} provide further pieces of this picture. As a minimal statistical framework for quantum chaotic behavior, RMT offers a natural starting point for identifying universal features of this interplay. Here, we address the fundamental question of whether the existence of a ``magic barrier'' during quantum thermalization can be established from RMT alone.

\section{Magic Barrier from RMT}

In order to study thermalization under random unitary time evolution we calculate the time dependence of the reduced density matrix in Gaussian RMT.
Then, to quantify quantum thermalization we investigate four different measures of complexity: entanglement entropy, antiflatness~\cite{Tirrito:2023fnw}, capacity of entanglement~\cite{Cao:2024nrx} and nonlocal trace distance magic (NLTDM)~\cite{Ebner:2026kwq,Sierant:2026qji,Viscardi:2026oos}.
Antiflatness and capacity represent witnesses of magic, while entanglement entropy provides a clock of thermalization. With NLTDM, we calculate a genuine magic measure during quantum thermalization.

\subsection{Operator expectation value}
To show how the magic barrier emerges from RMT, we first calculate the time dependent ensemble averaged expectation value of an arbitrary operator $P$ under the time evolution of a GOE Hamiltonian. For simplicity, this calculation will only be written out for a GOE Hamiltonian. For GUE/GSE, one uses unitary/symplectic Haar ensembles \cite{Collins_2009}, which results in a different four point correlator $C_{4pt}$, as compared to Eq.~(\ref{eq:four_point_corrlator}) and explanations above. However, the leading order result is the same.

For the entanglement measures discussed in this work, such as the antiflatness, one can show that the ensemble average commutes with the measure in the large $N$ limit, i.e., $\langle\AF(\rho_A)\rangle = \AF(\langle \rho_A \rangle)+\mathcal{O}(N^{-1})$, where $N$ is the Hilbert space dimension. Thus, it is possible to derive an analytic expression for these measures in the leading order of the $\frac{1}{N}$ expansion.

The following calculation closely follows the approach developed in \cite{Weidenmuller:2022wdc}. We consider a Hamiltonian $H$ drawn from the GOE, whose matrix elements $H_{\mu\nu}$ are real-valued Gaussian random variables with zero mean and variances $\langle H_{\mu\nu}^2\rangle=(1+\delta_{\mu\nu})\,\sigma^2/N$. The parameter $\sigma$ merely sets the energy scale and for large $N$ it is equal to the standard deviation of the eigenvalue spectrum, as specified in Sec.~\ref{s:goe_reduced_density_matrix}. The Hamiltonian admits the spectral decomposition $H_{\mu\nu} = \sum_{\alpha} O_{\mu\alpha} E_{\alpha} O_{\nu\alpha}$ where $E_{\alpha}$ are the eigenvalues of $H$, and $O_{\mu\alpha}$ are the components of the corresponding orthonormal eigenvectors. The matrix $O$ is a random orthogonal matrix distributed according to the Haar measure on the orthogonal group $O(N)$.

The matrix elements of the time evolution operator are given by $ U_{\mu\nu}(t) = \sum_{\alpha} O_{\mu\alpha} \exp\left(-iE_{\alpha}t\right) O_{\nu\alpha}$.

The time evolution of an arbitrary operator $P$ is given by
\begin{equation}
    \langle \mathrm{Tr}P\rho(t)\rangle = \left\langle\mathrm{Tr}\bigg( P \sum_{\alpha,\alpha^{\prime}}\Pi_{\nu\nu^{\prime}}e^{it(E_{\alpha^{\prime}}-E_{\alpha})}C_{4pt}\bigg)\right\rangle_{E_\alpha}
    \label{eq:expectation_value}
\end{equation}
because $P$ and $\Pi$ are not affected by the ensemble average. $\Pi$ is the initial density matrix of the system with $\mathrm{Tr}\Pi=1$. The average can be taken separately for the eigenfunctions $O_{\mu\nu}$ and the eigenvalues $E_{\alpha}$, as they are statistically independent \cite{Wettig:1999goe, Mehta:2004rmt}.
The required four point correlator $\langle U_{\mu\nu} U_{\mu^{\prime}\nu^{\prime}}^{\dagger}\rangle_O \sim \langle O_{\mu\alpha}O_{\nu\alpha}O_{\mu^{\prime}\alpha^{\prime}}O_{\nu^{\prime}\alpha^{\prime}}\rangle_O\equiv C_{4pt}$ for any finite dimension $N$ is given by \cite{Collins_2009}
\begin{widetext}
\begin{align}
\begin{split}
    C_{4pt}&=\frac{(N+1)}{N(N-1)(N+2)}\left[\delta_{\mu\nu}\delta_{\mu^{\prime}\nu^{\prime}}+\delta_{\alpha\alpha^{\prime}}(\delta_{\mu\mu^{\prime}}\delta_{\nu\nu^{\prime}}+\delta_{\mu\nu^{\prime}}\delta_{\mu^{\prime}\nu})\right]\\
    &-\frac{1}{N(N-1)(N+2)}\left[\delta_{\mu\nu}\delta_{\mu^{\prime}\nu^{\prime}}+\delta_{\mu\nu}\delta_{\mu^{\prime}\nu^{\prime}}+\delta_{\mu\mu^{\prime}}\delta_{\nu\nu^{\prime}}\delta_{\alpha\alpha^{\prime}}\right]\\
    &-\frac{1}{N(N-1)(N+2)}\left[\delta_{\mu\mu^{\prime}}\delta_{\nu\nu^{\prime}}+\delta_{\mu\nu^{\prime}}\delta_{\mu^{\prime}\nu}\delta_{\alpha\alpha^{\prime}}+\delta_{\mu\nu^{\prime}}\delta_{\mu^{\prime}\nu}\right].
    \label{eq:four_point_corrlator}
\end{split}
\end{align}
\end{widetext}
To leading order in $N$, the first coefficient is given by $\sim\frac{1}{N^2}+\order{N^{-4}}$ while the second coefficient is given by $\sim\frac{1}{N^3}+\order{N^{-4}}$. Since \cite{Weidenmuller:2022wdc} calculated the thermalization behavior in the large $N$ limit, only the leading order $\sim \frac{1}{N^2}$ was taken into account, i.e., all negative contributions from the second coefficient were omitted. However, because factors of $N$ arise during the summation process, terms with the second coefficient also contribute in the next to leading order $\order{N^{-1}}$. Thus, the final result in Eq.~(17) of \cite{Weidenmuller:2022wdc} is not the full $\order{N^{-1}}$ expansion, which results in $t=0$ being a singular point. This, however, does not weaken the main result of \cite{Weidenmuller:2022wdc} that the expectation value of any operator thermalizes in the $N\rightarrow \infty$ limit.

For our purposes this approximation is not sufficient to recover the the leading order physical behavior of the entanglement measures at any finite point in time. Thus, the following calculation will be performed using the full four point correlator of Eq.~(\ref{eq:four_point_corrlator}).
Inserting the correlator into Eq.~(\ref{eq:expectation_value}) and contracting all indices, one arrives at:

\begin{equation}
\begin{aligned}
    \delta_{\mu\nu}\delta_{\mu'\nu'}
        &\rightarrow \mathrm{Tr}(P\Pi),
        &\quad
    \delta_{\mu\nu'}\delta_{\mu'\nu}
        &\rightarrow \mathrm{Tr}(P),
        \\
    \delta_{\mu\mu'}\delta_{\nu\nu'}
        &\rightarrow \mathrm{Tr}(P^T\Pi),
        &
    \delta_{\alpha\alpha'}
        &\rightarrow N,\\
    1   &\rightarrow N^2\langle |f(t)|^2\rangle
\end{aligned}
\end{equation}
with $\mathrm{Tr}\Pi=1$ and $f(t)=\frac{1}{N}\sum_{\alpha}\exp{-i t E_{\alpha}}$. The quantity $K(t) = \langle|f(t)|^2\rangle$ is the well-known (normalized) spectral form factor \cite{Brezin:1997SFF,Das:2025sff}. The final result is given by
\begin{align}
\begin{split}
    \langle\mathrm{Tr(P\rho(t))}\rangle &= \frac{N(N+1)\langle|f(t)|^2\rangle-2}{(N-1)(N+2)}\mathrm{TrP\Pi}\\&+\frac{N[1-\langle|f(t)|^2\rangle]}{(N-1)(N+2)}\left[\mathrm{Tr}P+\mathrm{Tr}P^T\Pi\right].
    \label{eq:full_solution}
\end{split}
\end{align}
 Using $f(0)=1$, one can now show that the full finite $N$ solution has the correct $t=0$ limit:
\begin{equation}
    \frac{N(N+1)-2}{(N-1)(N+2)}=1 \rightarrow \langle\mathrm{Tr(P\rho(0))}\rangle = \mathrm{TrP\Pi}.
\end{equation}
Eq.~(\ref{eq:full_solution}) can now be expanded in orders of $\frac{1}{N}$, which reproduces Eq.~(11) of \cite{Weidenmuller:2022wdc} in the leading $\frac{1}{N}$ order with additional terms:
\begin{align}
\begin{split}
   \langle \mathrm{Tr}(P\rho(t))\rangle&= \frac{\mathrm{Tr P}}{N}\left(1-\frac{1}{N}\right) \left[1-\langle|f(t)|^2\rangle\right]\\\
   &+\mathrm{Tr}P\Pi \; \langle|f(t)|^2\rangle \\
   &+\frac{\mathrm{Tr}P^T\Pi}{N}\left[1-\langle|f(t)|^2\rangle\right]+\mathcal{O}(\frac{1}{N^2}).
   \label{eq:first_order_full_solution}
\end{split}
\end{align}
These additional terms now ensure that $\mathrm{Tr}(\rho(t))=1$ at any time, as can be tested by inserting $P = \mathds{1}$.

Performing this calculation with a GUE/GSE Hamiltonian results in the same leading order $\mathcal{O}(N^0)$ expression \cite{Weidenmuller:2022wdc} as given in Eq.~(\ref{eq:first_order_full_solution}). However, since both the Weingarten coefficients \cite{Collins_2009} and the delta contractions are different for GUE, one would expect the result to differ in all orders of $N^{-l}$ with $l\geq 1$. For GSE, since the symplectic calculation is similar to the othorgonal case, one requires the same Weingarten coefficients and index pairings. However, the symplectic structure modifies the index dependent terms in a non trivial way such that one expects all orders of $N^{-l}$ with $l\geq 1$ to differ from the othorgonal case.\footnote{For GUE, one does not obtain any $\mathrm{Tr}(P^T\Pi)$ terms. For GSE, pairings comparable to the ones that lead to $\mathrm{Tr}(P^T\Pi)$ in the GOE case do arise, however, they exhibit a significantly different structure.}

\subsection{Reduced density matrix}
\label{s:goe_reduced_density_matrix}
Let $A$ be a subsystem of the Hilbert space $\mathcal{H}=\mathcal{H}_A\otimes \mathcal{H}_{B}$. $\mathcal{H}_A$ is spanned by an orthogonal set of eigenfunctions $\ket{a},\ket{a^{\prime}}$. The reduced density matrix $\rho_A=\mathrm{Tr}_{B}\rho$ is obtained from the full density matrix $\rho(t)$ by projecting onto the corresponding $\mathcal{H}_A$:
\begin{align}
    \mathrm{Tr}_A(M\rho_A)&=\mathrm{Tr}((M\otimes \mathds{1})\rho)\label{eq:partial_trace}\\
\begin{split}
    \mathrm{Tr}((\ket{a}\bra{a^\prime}\otimes\mathds{1}_{B})\rho)&=\mathrm{Tr}_A(\ket{a}\bra{a^\prime}\rho_A)\\
    &=\sum_{a^{\prime\prime}}\bra{a^{\prime\prime}}\ket{a}\bra{a^{\prime}}\rho_A\ket{a^{\prime\prime}}=\bra{a^{\prime}}\rho_A\ket{a}.
\end{split}
\end{align}
Eq.~(\ref{eq:partial_trace}) is a well known identity for partial traces~\cite{Nielsen:2012yss}.
Thus, one can use the projector 
\begin{equation}
    P = \ket{a}\bra{a^\prime}\otimes\mathds{1}_{B}
\end{equation}
to calculate the ensemble averaged reduced density matrix for finite dimension $N$ from Eq.~(\ref{eq:full_solution}):
\begin{align}
\begin{split}
    \bra{a^{\prime}}\langle\rho_A\rangle\ket{a} &= \mathrm{Tr}\left((\ket{a}\bra{a^\prime}\otimes\mathds{1}_{B})\rho(t)\right)\\
    &=\frac{N(N+1)\langle|f(t)|^2\rangle-2}{(N-1)(N+2)}\bra{a^{\prime}}\mathrm{Tr}_{B}\Pi\ket{a}\\&+\frac{N[1-\langle|f(t)|^2\rangle]}{(N-1)(N+2)}\left[\delta_{a,a^{\prime}}N_{B}+\bra{a}\mathrm{Tr}_{B}\Pi\ket{a^{\prime}}\right]
    \end{split}
\end{align}
The first order in $N$ is obtained analogously by inserting the projector $P=\ket{a}\bra{a^\prime}\otimes\mathds{1}_{B}$ into Eq.~(\ref{eq:first_order_full_solution}).

Up to here, $\Pi$ was an arbitrary initial density matrix. For the following discussion, we now assume the initial state to be a real pure product state over the cut of the subsystems $A$ and $B$, which also results in $\Pi$ being rank one. This choice ensures the emergence of the magic barrier, since the initial state has no entanglement between the subsystems. W.l.o.g. we choose $\Pi = \ket{0}\bra{0}$. Since the GOE is invariant under orthogonal transformations and every real product state can be mapped onto $\ket{0}$ by a local orthogonal rotation, which leaves the entanglement spectrum unchanged, this is equivalent to an arbitrary real initial product state.

The leading order eigenvalues of the reduced density matrix are 
\begin{align}\label{eq:eigenvalues_of_RDM}
    \lambda_1=\frac{1+(m-1)\langle|f(t)|^2\rangle}{m},
    \quad
    \lambda_i=\frac{1-\langle|f(t)|^2\rangle}{m},
\end{align}
where $2\leq i\leq m$, i.e., there are only two distinct eigenvalues.\\
At fixed $t$ and for large $N$, the spectral form factor $K(t)=\langle|f(t)|^2\rangle$ is dominated by its disconnected part $|\langle f(t)\rangle|^2$, which is determined by the Fourier transform of the ensemble-averaged density of states. For all three Gaussian ensembles and large $N$, the latter is famously given by Wigner's semicircle law, $\rho_{\rm sc}(E)=\sqrt{4\sigma^2-E^2}/(2\pi\sigma^2)$ for $|E|\le 2\sigma$, where $\sigma$ is the standard deviation of the semicircle \cite{Mehta:2004rmt}. Finite-$N$ corrections to the mean density of states and the connected part, $\langle|f(t)|^2\rangle-|\langle f(t)\rangle|^2$, contribute at most with $\mathcal{O}(N^{-1})$, such that for all three Gaussian ensembles we get
\begin{align}
K(t) &= \Big|\int dE\,\rho_{\rm sc}(E)\,e^{-iEt}\Big|^2 + \mathcal{O}(N^{-1}) \nonumber\\
     &= \left[\frac{J_1(2\sigma t)}{\sigma t}\right]^2 + \mathcal{O}(N^{-1}),
\end{align}
where $J_1$ is the Bessel function of the first kind \cite{Mehta:2004rmt, Liu:2018sff, Das:2025sff}. Together with the identical leading-order coefficients of Eq.~(\ref{eq:first_order_full_solution}), this universality of $K(t)$ implies that all results below, especially the leading order expressions for the measures Eq.~(\ref{eq:EE})--(\ref{eq:capacity}) and Eq.~(\ref{eq:Mdist}), hold equally for GUE and GSE up to $\mathcal{O}(N^{-1})$ corrections.
\vspace*{1em}

\subsection{Entanglement and magic measures}\label{sec:analytical_measures}

In order to investigate the magic barrier we study four different measures of complexity: entanglement entropy, antiflatness, capacity of entanglement and nonlocal trace distance magic.
The entanglement entropy between the subsystem $A$ and its complement is defined as the von Neumann entropy
\begin{equation}\label{eq:EE_definition}
    S_A(\rho_A)=-\Tr(\rho_A\log\rho_A).
\end{equation}

The antiflatness of the entanglement spectrum is defined as~\cite{Tirrito:2023fnw}
\begin{equation}\label{eq:AF_definition}
    \AF(\rho_A)=\Tr_A\rho_A^3-(\Tr_A\rho_A^2)^2
\end{equation}
and measures how much the entanglement spectrum of the reduced density matrix deviates from a flat distribution.
It vanishes when the entanglement spectrum is flat, as e.g. for a maximally mixed or an initial product state. Further, the antiflatness has been shown to provide a lower bound on the nonlocal magic~\cite{Cao:2024nrx}, i.e. the nonlocal contributions to non-stabilizerness that cannot be removed by local unitaries.

Furthermore, capacity of entanglement for a pure quantum state is defined as~\cite{Cao:2024nrx}
\begin{equation}\label{eq:CA_definition}
    C_A(\rho_A)=T(\rho_A)-S_A(\rho_A)^2,
    \quad T(\rho_A)=\Tr(\rho_A(\log\rho_A)^2).
\end{equation}
It is the variance of the entanglement spectrum with respect to the reduced density matrix and a measure of antiflatness~\cite{Cao:2024nrx}. Hence, a nonzero capacity implies the presence of nonlocal magic~\cite{Cao:2024nrx}.
Capacity of entanglement has been explored in condensed matter systems~\cite{Yao:2010woi,Schliemann:2011zkg} as well as in the context of quantum gravity~\cite{DeBoer:2018kvc,Nakaguchi:2016zqi,Bueno:2022jbl}.

In Appendix~\ref{appendix:bounds} we show that the deviations between the averaged measures and the measures of the averaged reduced density matrix (RDM) are bounded by the mean squared fluctuations of the RDM. The latter is shown to be of order $\order{N^{-1}}$ as derived in Appendix~\ref{appendix:fluctuations}.
As a result, to leading order in large $N$, the averaged entanglement entropy, antiflatness, and capacity are given by Eq.~(\ref{eq:EE_definition})--(\ref{eq:CA_definition}) with $\rho_A=\langle\rho_A(t)\rangle$.
Given the eigenvalues of the averaged RDM in Eq.~(\ref{eq:eigenvalues_of_RDM}), the averaged measures read
\begin{widetext}
    \begin{align}
    \langle S_A(t)\rangle\label{eq:EE}
    &= \ln m
    -\frac{1+(m-1)K(t)}{m}\ln\!\bigl[1+(m-1)K(t)\bigr]
    -\frac{(m-1)[1-K(t)]}{m}\ln\!\bigl[1-K(t)\bigr]+\order{N^{-1}}
    \\\
    \langle\AF(t)\rangle
    &= \frac{m-1}{m^2}\,
    K(t)^2[1-K(t)][1+(m-1)K(t)]+\order{N^{-1}}\label{eq:anti_flatness}
    \\\
    \langle C_A(t)\rangle
    &= \frac{m-1}{m^2}\,
    [1-K(t)][1+(m-1)K(t)]
    \left[
    \ln\!\frac{1+(m-1)K(t)}{1-K(t)}
    \right]^2+\order{N^{-1}}.\label{eq:capacity}
    \end{align}
\end{widetext}

The maximum of the antiflatness $\langle \mathcal{F}(t)\rangle$ is reached at the time, which satisfies the condition
\begin{equation}
    K(t) = \frac{-6 + 3 m + \sqrt{4 - 4 m + 9 m^2}}{8 (m-1)}\xrightarrow{m\rightarrow\infty}\frac{3}{4}. 
    \label{eq:maximal_K}
\end{equation}
Remarkably, the corresponding eigenvalue spectrum in Eq.~(\ref{eq:eigenvalues_of_RDM}) then has exactly the same expression as the universal spectrum which maximizes the upper bound on the antiflatness for a general density matrix of any finite dimension $m$, as derived in Appendix J of Ref.~\cite{Jasser:2026hrg}. For large $m$ the maximal antiflatness is given by
\begin{equation}
     \langle\AF(t)\rangle_{\rm max}= \frac{27}{256}-\frac{9}{128m}+\mathcal{O}(m^{-2}),
\end{equation}
and the finite $m$ result is given in \cite{Jasser:2026hrg}. 

The fact that the bound is maximized during thermalization can be easily understood. The GOE is invariant under orthogonal transformations, so it singles out no direction in Hilbert space except the one set by the initial state. As a result, the RDM can contain only one distinguished eigenvalue, associated with the memory of the initial state, while all directions orthogonal to it are equivalent by symmetry. This forces the remaining $m-1$ eigenvalues to be degenerate. The RDM eigenvalue spectrum therefore has the required maximizing form (two distinct eigenvalues with multiplicities $1$ and $m-1$) at all times. It must then pass through the configuration (for large $m$) $\{\lambda_1,\lambda_{>1}\}=\{3/4,1/(4(m-1))\}$ during time evolution from the initial ($\{1,0\}$) to the final ($\{1/m,1/m\}$) configuration. In the lower panel of Fig.~\ref{fig:numerics_barrier} we show the time evolution of the two distinct eigenvalues.
Physical quantum chaotic Hamiltonians lack this full basis invariance because of locality, few-body interactions, and conservation laws. The resulting residual structure lifts the degeneracy of the $m-1$ eigenvalues, so the bound is approached closely but not reached exactly~\cite{Ebner:2025pdm}. Saturating the bound can therefore be read as a fingerprint of maximally structureless scrambling: the GOE realizes the extreme case that generic chaotic dynamics can only approximate.

So far we have only evaluated witnesses of magic, but not nonstabilizerness itself. NLTDM is defined as the minimum trace distance to the local-unitary orbit of $\mathrm{STAB}_0$~\cite{Cao:2024nrx}, i.e.
\begin{equation}
    M_{\mathrm{dist}}^{\mathrm{NL}}(\psi)
    =
    \min_{\substack{U=U_A\otimes U_B\\
    \sigma\in\mathrm{STAB}_0}}
    \frac12\|U\psi U^\dagger-\sigma\|_1.
\end{equation}
As shown in Appendix~\ref{appendix:NLTDM_bound_derivation} to leading order in large $N$ the averaged NLTDM is given by the NLTDM of the averaged RDM, i.e.,
\begin{align}
    \left|
    \langle M_{\mathrm{dist}}^{\mathrm{NL}}(\rho_A)\rangle
    -M_{\mathrm{dist}}^{\mathrm{NL}}(\langle\rho_A\rangle)
    \right|=\order{N^{-1/2}}.    
\end{align}
Hence, we identify the leading contribution to the NLTDM by calculating $M_{\mathrm{dist}}^{\mathrm{NL}}(\langle\rho_A\rangle)$.

So far we did not impose restrictions on the Hilbert space dimensionality. In the following, for the calculation of NLTDM, we restrict ourselves to qubit systems such that the Hilbert subspace of $A$ has dimension $m=2^k$, $k\geq1$.
For a pure bipartite state, the NLTDM is then given by~\cite{Ebner:2026kwq,Sierant:2026qji,Viscardi:2026oos}
\begin{align}\begin{split}
    M_{\mathrm{dist}}^{\mathrm{NL}}
    &=
    \sqrt{1-\max_{\Gamma\in\{1,2,4,\ldots,m\}}F_{\Gamma}},
    \\\
    F_{\Gamma}&=\frac{1}{\Gamma}
    \left(\sum_{i=1}^{\Gamma}\sqrt{\lambda_i}\right)^2,
\end{split}\end{align}
where $\{\lambda_i\}$ denotes the decreasingly ordered set of eigenvalues of $\rho_A$. 
Let $a=\lambda_1$ and $b=\lambda_{i>1}$ given by Eq.~(\ref{eq:eigenvalues_of_RDM}). We have $a\geq b$ and thus
\begin{align}
    F_{\Gamma}
    &=\frac{\bigl[\sqrt a+(\Gamma-1)\sqrt b\bigr]^2}{\Gamma}
\end{align}
is convex. As a consequence its maximum on the finite interval $[1,m]$ lies at an endpoint, i.e.,
\begin{align}
    \max_{\Gamma} F_{\Gamma}
    =
    \max\left\{a,\,\frac{\bigl[\sqrt a+(m-1)\sqrt b\bigr]^2}{m}\right\}.
\end{align}
Thus, the leading order contribution of the averaged NLTDM reads
\begin{widetext}
    \begin{align}
        \langle M_{\mathrm{dist}}^{\mathrm{NL}}(\rho_A)\rangle
        =
        \min\left[\sqrt{\frac{(m-1)(1-K(t))}{m}},\frac{\sqrt{m-1}}{m}\left(\sqrt{1+(m-1)K(t)}-\sqrt{1-K(t)}\right)\right]
        +\order{N^{-1/2}}.\label{eq:Mdist}
    \end{align}
\end{widetext}
It is interesting to note that there is a subspace dimension dependent crossover value at
\begin{equation}
    K(t)=\frac{\sqrt{m}+2}{2(\sqrt{m}+1)}\xrightarrow{m\rightarrow\infty}\frac{1}{2}.
\end{equation}

The upper panel of Fig.~\ref{fig:numerics_barrier} shows the time evolution of the four quantum information measures analytically derived above. The quantum state evolves from the initially non-entangled product state to the maximally mixed state. As a result, the entanglement entropy increases from $0$ to the maximal value $\log(m)$.
On the other hand, the antiflatness, capacity and NLTDM vanish both for the initial product and final maximally mixed state. During the equilibration process all three exhibit a pronounced transient peak at a similar time scale given by the time of largest entanglement growth.
This establishes the magic barrier~\cite{Ebner:2025pdm} as a universal feature of thermalization in RMT.

Further, in Fig.~\ref{fig:numerics_scaling} we show the behavior of the corrections at finite total Hilbert space dimension $N$ using the numerical simulations described below.

\section{Numerical comparison}\label{sec:numerics}

We test the analytical results against numerics using exact diagonalization and averaging over many independent GOE matrix samples. As initial state we always use $\ket{\psi_0}=\ket{0}^{\otimes L}$, as discussed in Sec.~\ref{s:goe_reduced_density_matrix}.

Consider a subsystem $A$ consisting of $L_A=3$ qubits in the middle of a $L=12$ chain. The subsystem and total Hilbert space dimensions are $m=2^{L_A}=8$ and  $N=2^L=4096$.
From $\ket{\psi(t)}$ we obtain the reduced density matrix $\rho_A(t)$ of $A$.
The GOE averages of the quantum information measures are calculated by evaluating the measures for every Hamiltonian realization and averaging them, i.e. $\langle O(\rho_A)\rangle$.

\begin{figure}[t]
    \centering
    \includegraphics[width=\columnwidth]{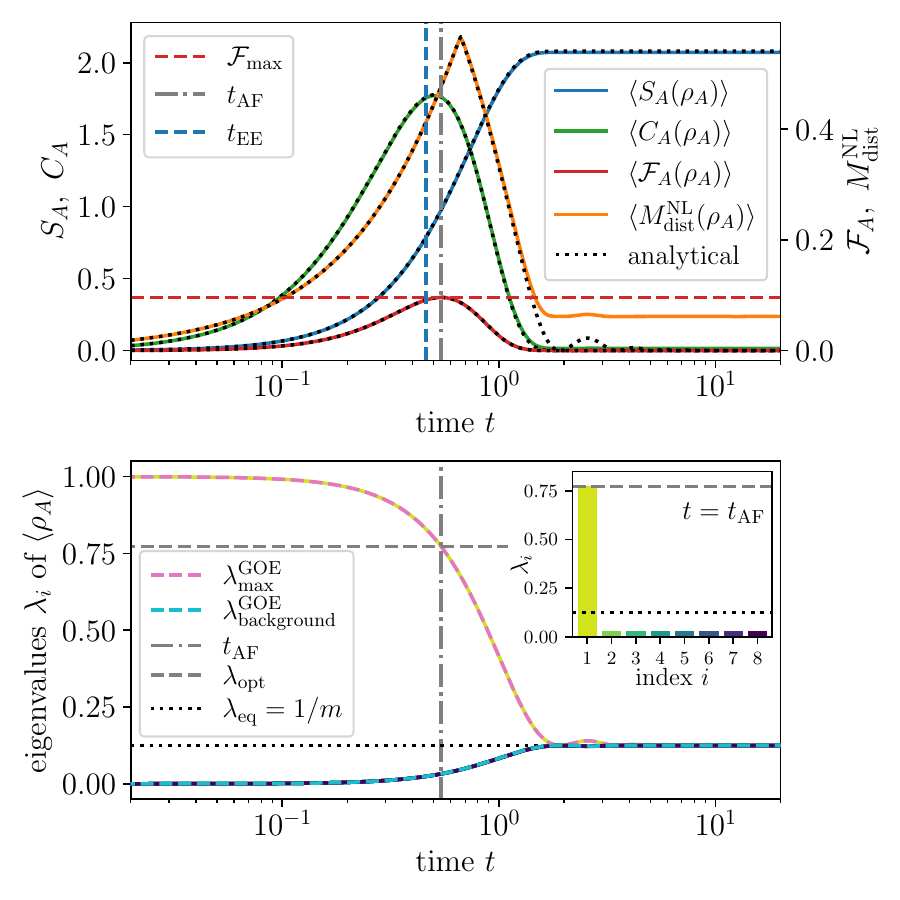}
    \caption{
    Comparison of analytical results with the numerical ensemble average for the time evolution under a GOE Hamiltonian for $L=12$ qubits ($N=2^{L}$), averaged over $1000$ samples, with $\sigma=1$, initial state $\ket{0}^{\otimes L}$, and subsystem $A$ of $L_A=3$ qubits in the middle of the chain ($m=2^{L_A}=8$).\\
    Top: numerical dynamics of the ensemble averaged entanglement entropy $\langle S_A(\rho_A)\rangle$ (blue), capacity $\langle C_A(\rho_A)\rangle$ (green), antiflatness $\langle \AF(\rho_A)\rangle$ (red), and NLTDM $\langle M_{\mathrm{dist}}^{\mathrm{NL}}(\rho_A)\rangle$ (orange) compared to their corresponding analytical large $N$ results (black dotted) obtained in Sec.~\ref{sec:analytical_measures}. The vertical line $t_{EE}$ (blue dashed) marks the time at which the entanglement entropy grows most rapidly, i.e. the maximum of $dS_A/dt$, and $t_{\rm AF}$ (gray dash-dotted) the time of maximal antiflatness.\\
    Bottom: eigenvalues $\lambda_i$ of $\langle\rho_A\rangle$ together with the analytical results for the largest RDM eigenvalue $\lambda_\mathrm{max}^\mathrm{GOE}$ (pink dashed) and $(m\!-\!1)$-fold degenerate background $\lambda_\mathrm{background}^\mathrm{GOE}$ (cyan dashed) of Eq.~(\ref{eq:eigenvalues_of_RDM}) , and the equilibrium value $\lambda_{\rm eq}=1/m$ (dotted). Inset: eigenvalues $\lambda_i$ at $t=t_{\rm AF}$. The horizontal lines indicate the analytical maximal antiflatness attainable for a density matrix with dimension $m=8$, $\langle\AF\rangle_{\max}\approx0.096$ (red dashed, top), and the optimizing eigenvalue at which it is attained, $\lambda_{\rm opt}\approx0.772$ (gray dashed, bottom). For $m\to\infty$ they approach $27/256\approx0.105$ and $3/4$, as shown in Appendix J of Ref.~\cite{Jasser:2026hrg}. The standard error for all quantities is not shown as it is smaller than linewidth.}
    \label{fig:numerics_barrier}
\end{figure}

Figure~\ref{fig:numerics_barrier} compares the numerical dynamics of entanglement entropy $\langle S_A(\rho_A)\rangle$, capacity $\langle C_A(\rho_A)\rangle$, antiflatness $\langle \AF(\rho_A)\rangle$, and NLTDM $\langle M_{\mathrm{dist}}^{\mathrm{NL}}(\rho_A)\rangle$ with the analytical large $N$ results obtained in Sec.~\ref{sec:analytical_measures}.
The antiflatness, capacity and magic vanish for the initial product state, grow to a pronounced transient maximum, the magic barrier, and decay again as $\rho_A$ approaches the maximally mixed state. At the same time the entanglement entropy grows from $0$ to $\log(m)$.
The numerical data closely follows the analytical results for the entanglement entropy, antiflatness and capacity. We find the NLTDM at late times to not vanish as predicted by our analytical formula, but equilibrate to a small finite value. We attribute this to finite size effects, since for $N=4096$ the $\order{N^{-1/2}}$ corrections to the analytical result are of the same order as the equilibration value.

The time at which the antiflatness reaches its maximum, is closely correlated with the time of fastest entanglement growth, i.e.\ the maximum of $dS_A/dt$, similar to what was observed in \cite{Ebner:2025pdm,Zhang:2026ucb} for physical systems.
The lower panel shows the origin of the barrier: the spectrum of $\langle\rho_A\rangle$ consists of one large eigenvalue and an $(m-1)$-fold degenerate background, as predicted by Eq.~(\ref{eq:eigenvalues_of_RDM}), and the antiflatness is maximal when the largest eigenvalue passes through the optimizing eigenvalue $\lambda_{\rm opt}$, that maximizes $\AF$ for such a spectrum \cite{Jasser:2026hrg}.
Similarly, for the maxima of the capacity and NLTDM we find that their times do not agree exactly, but lie in the same time regime, i.e. where entanglement entropy grows rapidly.

\begin{figure}[t]
    \centering
    \includegraphics[width=\columnwidth]{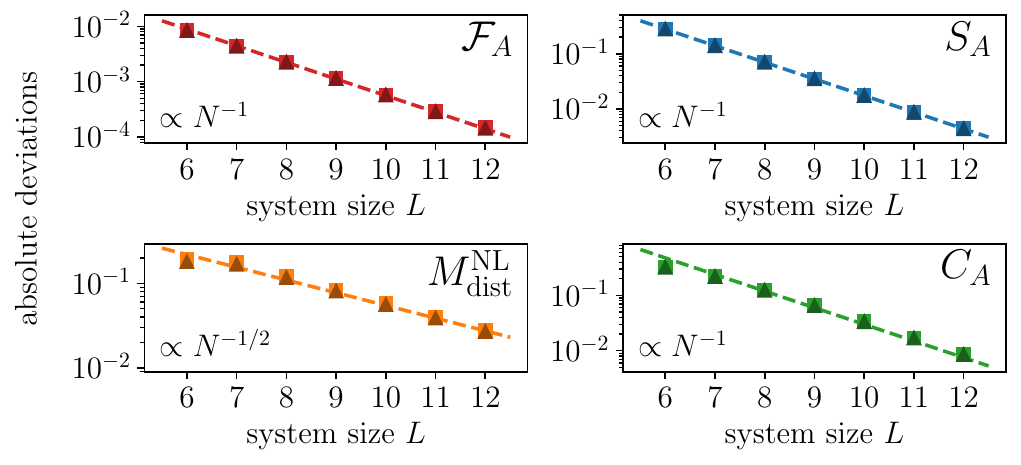}
    \caption{Comparison of the finite-size scaling for the numerical ensemble averaged entanglement entropy $S_A$ (blue), capacity $C_A$ (green), antiflatness $\AF$ (red), and NLTDM $M_{\mathrm{dist}}^{\mathrm{NL}}$ (orange), for different system sizes ($N=2^L$), $1000$ GOE samples per $L$, $\sigma=1$, initial state $\ket{0}^{\otimes L}$, and subsystem $A$ of $L_A=3$ qubits ($m=8$) in the middle of the chain.
    Shown is the absolute deviation between the two averaging orders $|\langle O(\rho_A)\rangle-O(\langle\rho_A\rangle)|$ (square markers) and the absolute deviation from their analytical large N results $|\langle O(\rho_A)\rangle- \langle O \rangle_\mathrm{analytic}|$ (triangle markers, darker color) in Sec.~\ref{sec:analytical_measures} for the four quantities $O\in \{S_A ,\, C_A ,\, \AF ,\, M_{\mathrm{dist}}^{\mathrm{NL}}\}$. The absolute deviation was time averaged over 400 logarithmically spaced points between $t\!=\!10^{-2}$ and $10^2$. The standard error for all quantities is not shown as it is smaller than linewidth. Dashed lines are the $\propto N^{-1}$ (and $\propto N^{-1/2}$ for NLTDM) fits for the difference between the two averaging orders, which all show the expected scaling behavior.}
    \label{fig:numerics_scaling}
\end{figure}

Figure~\ref{fig:numerics_scaling} quantifies the finite-size corrections of the four quantum information measures, i.e. $|\langle O(\rho_A)\rangle-O(\langle\rho_A\rangle)|$. As derived in Appendix~\ref{appendix:bounds}, the deviations are bounded by $\order{N^{-1}}$ for $S_A$, $\AF$, and $C_A$, where by $\order{N^{-1/2}}$ for the NLTDM.
This confirms numerically the analytical large $N$ formulas, Eq.~(\ref{eq:EE})--(\ref{eq:capacity}) and Eq.~(\ref{eq:Mdist}), as well as the order of corrections.

\section{Conclusion}
We investigated the time evolution of four measures of complexity, namely the entanglement entropy, antiflatness, capacity of entanglement and the nonlocal trace distance magic (NLTDM), during thermalization under a Gaussian random Hamiltonian in the large $N$ limit using random matrix theory.
To obtain analytical expressions for these quantities, we first calculated the ensemble averaged reduced density matrix $\langle \rho_A (t)\rangle$ for all dimensions $N$ by explicitly evaluating the time evolution under the random Hamiltonian. We show that in the large $N$ limit, the complexity measures in question commute with the ensemble average up to $\mathcal{O}(N^{-1})$ ($\mathcal{O}(N^{-1/2})$ for the NLTDM) corrections, i.e., $\langle O(\rho_A)\rangle=O(\langle \rho_A\rangle)+\mathcal{O}(N^{-1})$, which allows us to obtain their leading order large $N$ limit by using the previously derived averaged RDM.
The antiflatness, capacity and NLTDM show a transient maximum, i.e. a magic barrier, in the time regime of maximal entanglement growth. The leading order results for all three Gaussian ensembles are identical.
We verified the analytical results as well as the derived correction order with numerical simulations of GOE random unitary time evolution.
We conclude that the magic barrier is a universal feature of ergodic quantum theories obeying RMT level statistics and that quantitative simulations of thermalization in these theories inherently require quantum computation.
Future direction involve investigating certain relaxations of our model, e.g. by only requiring RMT locally or introducing an additional energy scale via the ETH.

\begin{acknowledgments}
The authors gratefully acknowledge valuable discussion with and useful comments on the manuscript from Lukas Ebner, Berndt M\"uller, and Xiaojun Yao.
A.S., J.M., and C.S. are supported by the DFG (German Research Foundation, grant 553079183).
C.S. acknowledges financial support from the German Academic Scholarship Foundation (Studienstiftung des deutschen Volkes) and the Elite Network of Bavaria through the Marianne-Plehn-Program.
We acknowledge the use of OpenAI's GPT-6 Astra to assist in the derivation of mathematical bounds for Sections~\ref{appendix:Capacity_bound_derivation} and \ref{appendix:NLTDM_bound_derivation} as well as for independent consistency checks in Sec.~\ref{appendix:fluctuations}.
All analytical results were independently derived and verified by the authors.
\end{acknowledgments}

\appendix

\section{Mean squared fluctuations of the RDM}\label{appendix:fluctuations}
Let $\Delta=\rho_A-\langle\rho_A\rangle$ denote the difference between the RDM and its average with $\langle \Delta \rangle = 0$.
Our goal is to show that the mean squared fluctuations $\langle\Tr\Delta^2\rangle$ scale as $\order{N^{-1}}$.
We introduced the $m^2$-dimensional orthonormal basis of hermitian $m\times m$ matrices
\begin{equation}
    B_0=\frac{\mathds{1}_m}{\sqrt{m}},
    \quad
    B_1,\dots,B_{m^2-1},
    \quad
    \Tr_A(B_aB_b)=\delta_{ab}
\end{equation}
where $\Tr B_a=0$ for $a>0$. Since $\Tr_A\Delta=0$ we can expand
\begin{equation}
    \Delta=\sum_{a=1}^{m^2-1}d_aB_a,\quad d_a=\Tr_A(B_a\Delta).
\end{equation}
The $d_a$ are real since $B_a$ and $\Delta$ are Hermitian. It is easy to check that
\begin{equation}
    \Tr_A\Delta^2=\sum_{a=1}^{m^2-1}d_a^2.
\end{equation}
Defining $x_a=\Tr_A(B_a\rho_A)$ one can check that $d_a=x_a-\langle x_a\rangle$. As a result the mean squared fluctuations of the density matrix can be written as
\begin{align}
    \langle\Tr_A\Delta^2\rangle
    =\sum_{a=1}^{m^2-1}\langle x_a^2\rangle-\langle x_a\rangle^2.
\end{align}
For convenience define the full Hilbert space matrix $\mathcal{A}_a=\mathds{1}_N+B_a\otimes \mathds{1}_n$ since $\Tr_{AB}(\mathcal{A}_a\rho)=1+x_a$ and $\langle x_a^2\rangle-\langle x_a\rangle^2=\langle (1+x_a)^2\rangle-\langle 1+x_a\rangle^2$ such that the fluctuation average can be written as
\begin{align}
    \langle\Tr_A\Delta^2\rangle
    =\sum_{a=1}^{m^2-1}\langle (\Tr_{AB}(\mathcal{A}_a\rho))^2\rangle-\langle \Tr_{AB}(\mathcal{A}_a\rho)\rangle^2
\end{align}
The density matrix corresponds to a rank-one initial state and we can verify that
\begin{align}
    \frac{\Tr_{AB}\mathcal{A}_a^2}{(\Tr_{AB}\mathcal{A}_a)^2}=\frac{N+n}{N^2}=\order{N^{-1}}.
\end{align}
Further, the operator norm of $\mathcal{A}_a$ is bounded in $N$. It follows that~\cite{Weidenmuller:2022wdc}
\begin{align}
    \langle (\Tr_{AB}(\mathcal{A}_a\rho))^2\rangle-\langle \Tr_{AB}(\mathcal{A}_a\rho)\rangle^2
    =\order{N^{-1}}.
\end{align}
Consequently, the mean square of fluctuations of the reduced density matrix
\begin{align}
    \langle\Tr_A\Delta^2\rangle
    =\sum_{a=1}^{m^2-1}\order{N^{-1}}=\order{N^{-1}}
\end{align}
scales as expected and vanishes in the large $N$ limit.

\section{Bounds on entanglement and magic measures}\label{appendix:bounds}

\subsection{Entanglement entropy}\label{appendix:EE_bound_derivation}
Realize that
\begin{align}
    S(\langle\rho_A\rangle)-\langle S(\rho_A)\rangle
    &=
    \langle D(\rho_A||\langle\rho_A\rangle)\rangle,
    \label{eq:entropy_difference_from_relative}
\end{align}
i.e. the difference between the averaged entanglement entropy and the entanglement entropy of the averaged reduced density matrix is controlled by the averaged relative entropy
\begin{equation}
    D(\rho_A||\langle\rho_A\rangle)
    =\Tr\left(\rho_A(\log\rho_A-\log \langle\rho_A\rangle)\right).
\end{equation}
For positive definite $\langle\rho_A\rangle$ the relative entropy is bounded by~\cite{Audenaert:2005pct}
\begin{equation}
    0\leq D(\rho_A||\langle\rho_A\rangle)
    \leq\frac{\Tr(\rho_A-\langle\rho_A\rangle)^2}{\lambda_{\rm min}(\langle\rho_A\rangle)},
\end{equation}
where $\lambda_{\rm min}(\langle\rho_A\rangle)$ denotes the smallest eigenvalue of $\langle\rho_A\rangle$. Hence, we can estimate the difference between $\langle S_A(\rho_A)\rangle$ and $S_A(\langle\rho_A\rangle)$ as
\begin{equation}
    0\leq S_A(\langle\rho_A\rangle)-\langle S_A(\rho_A)\rangle\leq
    \frac{\langle\Tr\Delta^2\rangle}{\lambda_{\rm min}(\langle\rho_A\rangle)}
    =\order{N^{-1}},
\end{equation}
where we used the result of Appendix~\ref{appendix:fluctuations}. Furthermore, for fixed $m>1$ and fixed $t>0$, the smallest eigenvalue of $\langle \rho_A \rangle$ satisfies
\begin{align}
    \lambda_{\rm min}(\langle\rho_A\rangle)=\frac{1-\langle|f(t)|^2\rangle}{m}+\order{N^{-1}}\geq c>0
\end{align}
It is, therefore, bounded away from zero uniformly in $N$ for sufficiently large $N$. At $t=0$, both entropies vanish and the claimed equality holds exactly.
As a result, in leading order the averaged entanglement entropy is given by the entanglement entropy of the averaged reduced density matrix.

\subsection{Antiflatness}\label{appendix:AF_bound_derivation}
One can write the difference between the averaged antiflatness and antiflatness of the averaged RDM as
\begin{align}
    \begin{split}
        \langle\AF(\rho_A)\rangle-\AF(\langle\rho_A\rangle)
        =&
        \langle\Tr((\rho_A+2\langle\rho_A\rangle)\Delta^2)\rangle
        \\\
        &-2\Tr \langle\rho_A\rangle^2\langle\Tr\Delta^2\rangle
        \\\
        &-\langle(\Tr\rho_A^2-\Tr \langle\rho_A\rangle^2)^2\rangle
    \end{split}
\end{align}
Since $0\leq\rho_A,\langle\rho_A\rangle\leq \mathds{1}_m$ and $\Tr \langle\rho_A\rangle^2\leq1$ one can estimate
\begin{equation}
    0\leq\langle\Tr((\rho_A+2\langle\rho_A\rangle)\Delta^2)\rangle
    \leq3\langle\Tr\Delta^2\rangle.
\end{equation}
Further, using the trace norm and the Cauchy-Schwarz inequality, one finds
\begin{equation}
    (\Tr\rho_A^2-\Tr \langle\rho_A\rangle^2)^2\leq4\Tr\Delta^2
\end{equation}
The antiflatness difference then is thus bounded by
\begin{equation}
    |\langle\AF(\rho_A)\rangle-\AF(\langle\rho_A\rangle)|
    \leq6\langle\Tr\Delta^2\rangle=\order{N^{-1}},
\end{equation}
where we again used the result from Appendix~\ref{appendix:fluctuations}.
Thus, the averaged antiflatness is given by the antiflatness of the averaged reduced density matrix to leading order in large $N$.

\subsection{Capacity of entanglement}\label{appendix:Capacity_bound_derivation}
First, using Eq.~(\ref{eq:entropy_difference_from_relative}) one obtains
\begin{align}\label{eq:temp1}
    &\langle S_A(\rho_A)^2\rangle-S_A(\langle\rho_A\rangle)^2\\\
    &=\langle(S_A(\rho_A)-S_A(\langle\rho_A\rangle))^2\rangle
    -2S_A(\langle\rho_A\rangle)\langle D(\rho_A||\langle\rho_A\rangle)\rangle
    \nonumber
\end{align}
The second term is at most $\order{N^{-1}}$ since $0\leq S_A(\langle\rho_A\rangle)\leq\log(m)$ and $0\leq \langle D(\rho_A||\langle\rho_A\rangle)\rangle\leq\order{N^{-1}}$. The first term needs more care. For $t\neq0$ let $\lambda_{\rm min}(\langle\rho_A\rangle)\geq c>0$ where c is independent of $N$.
From the definition of relative entropy one obtains
\begin{equation}
    S_A(\rho_A)-S_A(\langle\rho_A\rangle)
    =-\Tr(\Delta\log\langle\rho_A\rangle)-D(\rho_A||\langle\rho_A\rangle)
\end{equation}
and hence, using $(a+b)^2\leq2a^2+2b^2$,
\begin{align}\nonumber
    &(S_A(\rho_A)-S_A(\langle\rho_A\rangle))^2
    \\\
    &\leq2(\Tr(\Delta\log \langle\rho_A\rangle))^2+2D(\rho_A||\langle\rho_A\rangle)^2
\end{align}
Using the Cauchy-Schwarz inequality, the upper bound
\begin{equation}
    \Tr(\log \langle\rho_A\rangle)^2=\sum_{j=1}^m(\log \lambda_j)^2\leq m(\log(1/c))^2,
\end{equation}
as well as $\Tr\Delta^2\leq2$ one finds
\begin{equation}
    \langle (S_A(\rho_A)-S_A(\langle\rho_A\rangle))^2 \rangle
    \leq
    \left(2m(\log(1/c))^2+\frac{4}{c^2}\right)\langle\Tr\Delta^2\rangle
\end{equation}
which is of order $\order{N^{-1}}$. Therefore, Eq.~(\ref{eq:temp1}) is of order $\order{N^{-1}}$. The remaining task is to evaluate the term involving $T(\rho)$.

We want to show that
\begin{equation}
    -\frac{2\log(1/c)}{c}\langle\Tr\Delta^2\rangle
    \leq \langle T(\rho_A)\rangle-T(\langle\rho_A\rangle)
    \leq \langle\Tr\Delta^2\rangle.
    \label{eq:T-two-sided-bound}
\end{equation}
To obtain the upper bound, consider the continuous function $h(x)=x(\log x)^2-x^2$, with $h(0)=0$. Since
\begin{equation}
    h''(x)=\frac{2}{x}\bigl(\log x+1-x\bigr)\leq0,
\end{equation}
$h$ is concave. Jensen's trace inequality \cite{Hansen:2003} implies
\begin{equation}
    \left\langle\operatorname{Tr}h(\rho_A)\right\rangle
    \leq \operatorname{Tr}h(\langle\rho_A\rangle).
\end{equation}
Consequently,
\begin{align}
    \langle T(\rho_A)\rangle-T(\langle\rho_A\rangle)
    &\leq
    \left\langle\operatorname{Tr}\rho_A^2\right\rangle
    -\operatorname{Tr}\langle\rho_A\rangle^2
    =\langle\Tr\Delta^2\rangle,
    \label{eq:T-upper-bound}
\end{align}
where we also used $\langle\Delta\rangle=0$.

For the lower bound we first define
\begin{equation}
    M=\langle\rho_A\log\rho_A\rangle-\langle\rho_A\rangle\log \langle\rho_A\rangle.
\end{equation}
Operator convexity of $x\log x$, together with operator Jensen's inequality, gives $M\geq0$~\cite{Hansen:2003}.
Moreover,
\begin{equation}
    \operatorname{Tr}M
    =\left\langle D(\rho_A\Vert \langle\rho_A\rangle)\right\rangle.
\end{equation}
and thus the bound on the relative entropy yields
\begin{equation}
    0\leq\operatorname{Tr}M
    \leq
    \frac{\langle\operatorname{Tr}\Delta^2\rangle}
         {c}=\order{N^{-1}}.
\end{equation}
Next, introduce the nonnegative quantity
\begin{equation}
    Q=
    \left\langle
    \operatorname{Tr}\!\left[
        \rho_A(\log\rho_A-\log \langle\rho_A\rangle)^2
    \right]
    \right\rangle
    \geq0,
\end{equation}
expand the square inside the trace and use the definition of $M$:
\begin{align}
    Q
    &=\langle T(\rho_A)\rangle-T(\langle\rho_A\rangle)
      -2\operatorname{Tr}(M\log \langle\rho_A\rangle).
\end{align}
Since $\log \langle\rho_A\rangle\geq(\log c)\mathds{1}_m$ and $M\geq0$, it follows that
\begin{align}
    \langle T(\rho_A)\rangle-T(\langle\rho_A\rangle)
    &\geq 2\operatorname{Tr}(M\log \langle\rho_A\rangle)\geq 2\log c\,\operatorname{Tr}M \nonumber\\
    &\geq -\frac{2\log(1/c)}{c}\langle\operatorname{Tr}\Delta^2\rangle.
\end{align}
Together with Eq.~\eqref{eq:T-upper-bound}, this proves
Eq.~\eqref{eq:T-two-sided-bound} and thus:
\begin{align}\nonumber
    &|\langle C_A(\rho_A)\rangle-C_A(\langle\rho_A\rangle)|
    \\\
    &\leq\nonumber
    |\langle T(\rho_A)\rangle-T(\langle\rho_A\rangle)|
    +|\langle S_A(\rho_A)^2\rangle-S_A(\langle\rho_A\rangle)^2|
    \\\
    &=\order{N^{-1}}
\end{align}

\subsection{Nonlocal trace distance magic}\label{appendix:NLTDM_bound_derivation}
The NLTDM fulfills the usual triangle inequality for distance-based entanglement measures~\cite{Tendick:2022wlg,Cao:2024nrx}
\begin{align}
    \left|
    M_{\mathrm{dist}}^{\mathrm{NL}}(\psi)
    -M_{\mathrm{dist}}^{\mathrm{NL}}(\phi)
    \right|
    \le\frac12\||\psi\rangle\langle\psi|-|\phi\rangle\langle\phi|\|_1.
\end{align}
To apply this inequality, choose the normalized purifications
\begin{equation}
|\Psi_X\rangle
=(\sqrt X\otimes \mathds{1}_B)\sum_{i=1}^{m}|i\rangle_A|i\rangle_B,
\quad X\in\{\rho_A,\langle\rho_A\rangle\}.
\end{equation}
Their overlap is $z=\langle\Psi_{\rho_A}|\Psi_{\langle \rho_A \rangle}\rangle=\Tr(\sqrt{\rho_A}\sqrt{\langle\rho_A\rangle})\in[0,1]$. For two pure states the trace distance is given by~\cite{Watrous:2018zil}
\begin{equation}
    \frac12\||\psi\rangle\langle\psi|-|\phi\rangle\langle\phi|\|_1
    =\sqrt{1-z^2}
\end{equation}
On the other hand
\begin{align}\nonumber
    \|\sqrt{\rho_A}-\sqrt{\langle\rho_A\rangle}\|_{HS}^2
    &=\operatorname{Tr}\rho_A+\operatorname{Tr}\langle\rho_A\rangle
    -2\operatorname{Tr}(\sqrt{\rho_A}\sqrt{\langle\rho_A\rangle})\\
    &=2(1-z).
\end{align}
Since $1-z^2\le2(1-z)$ we obtain
\begin{align}
    \left|
    M_{\mathrm{dist}}^{\mathrm{NL}}(\rho_A)
    -M_{\mathrm{dist}}^{\mathrm{NL}}(\langle\rho_A\rangle)
    \right|
    \le
    \|\sqrt{\rho_A}-\sqrt{\langle\rho_A\rangle}\|_{HS}
\end{align}
We write $M_{\mathrm{dist}}^{\mathrm{NL}}(\rho_A)$ for the nonlocal trace distance magic of a pure bipartite state with reduced density matrix $\rho_A$. As before using $\langle\rho_A\rangle\ge c\,\mathds{1}_m$ and defining $D=\sqrt{\rho_A}-\sqrt{\langle\rho_A\rangle}$ we find $\Delta=\sqrt{\rho_A}\,D+D\sqrt{\langle \rho_A \rangle}$ and thus
\begin{align}
    \operatorname{Tr}(D\Delta)
    &=\operatorname{Tr}(\sqrt{\rho_A}D^2)
    +\operatorname{Tr}(\sqrt{\langle\rho_A\rangle}\,D^2)
    \ge\sqrt c\,\operatorname{Tr}D^2.
\end{align}
which together with the Cauchy-Schwarz inequality yields
\begin{align}
    \left|
    \langle M_{\mathrm{dist}}^{\mathrm{NL}}(\rho_A)\rangle
    -M_{\mathrm{dist}}^{\mathrm{NL}}(\langle \rho_A \rangle)
    \right|
    \leq
    \sqrt{\frac{\langle\Tr\Delta^2\rangle}{c}}=\order{N^{-1/2}}.
\end{align}
Thus, we have shown that to leading order in $N$ the average over the NLTDM is given by 
$M_{\mathrm{dist}}^{\mathrm{NL}}(\langle \rho_A \rangle)$. In contrast to the other measures investigated so far the deviations in the nonlocal trace distance magic are only bounded to order $\order{N^{-1/2}}$ here.

\newpage
\bibliographystyle{apsrev4-1}
\bibliography{biblio}

\end{document}